\documentclass{emulateapj}
\usepackage{amssymb,amsmath,graphicx,longtable,subfigure,wrapfig}
\usepackage{rotating}
\usepackage[colorlinks,linkcolor=blue,anchorcolor=green,citecolor=blue]{hyperref}
\usepackage{natbib}

\shorttitle{Low frequency view of NGC\,4869}
\shortauthors{Lal, D.V.}

\begin{document}

\title{NGC\,4869 in the Coma cluster: Twist, wrap, overlap, and bend}

\author{Dharam V. Lal}
\affil{National Centre for Radio Astrophysics - Tata Institute of Fundamental Research
 Post Box 3, Ganeshkhind P.O., Pune 41007, India}
\email{dharam@ncra.tifr.res.in}

\begin{abstract}

The upgraded Giant Metrewave Radio Telescope (GMRT) has been used to image the head-tail radio galaxy NGC\,4869 in the Coma cluster with an angular resolution of 6\farcs26 at 250--500 MHz and 2\farcs18 at the 1050--1450 MHz bands.  The archival legacy GMRT data have also been used to image the source with angular
resolutions from 4\farcs9 to 21\farcs8 at 610 MHz, 325 MHz, 240 MHz, and 150 MHz.  We find that the $\sim$200~kpc~scale radio morphology consists of five distinct regions with the clear presence of a pinch at $\approx$1\farcm4 (= 38.8~kpc) and a ridge at $\approx$3\farcm4 (= 94.2~kpc) from the head. The sharp bend by $\sim$70$^\circ$ at $\sim$3\farcm5 (= 97~kpc) from the head is possibly due to projection effects.  The radio spectra show progressive spectral steepening as a function of distance from the head and there is possibly re-acceleration of the synchrotron electrons and perhaps also magnetic field re-generation in the 6$^{\prime\prime}$--208$^{\prime\prime}$ (= 2.8--96.1~kpc) region of the jet.  We report a steep spectrum sheath layer enveloping a flat spectrum spine, hinting at a transverse velocity structure with a fast-moving spine surrounded by a slow-moving sheath layer. 
We also derive the lifetimes of the radiating electrons and equipartition parameters.
A plausible explanation for the characteristic feature, a ridge of emission perpendicular to the direction of tail is the flaring of a straight, collimated radio jet as it crosses a surface brightness edge due to Kelvin-Helmholtz instabilities.

\end{abstract}


\section{Introduction}
\label{intro}

Complex, unusual structures in radio sources are often associated with several events in which radio-emitting blobs of plasma have been created; for example, when the radio galaxy is moving through the intracluster medium (ICM), as is the case with the head-tail radio sources, a history of activity is traced in the tail of radio-emitting plasma left behind the parent host galaxy.  Within this picture, therefore, tailed sources are viewed as trails or fossil records deposited by active galaxies.  These sources are found in both poor groups and rich clusters of galaxies and their radio morphologies are due to the interaction of the radio-emitting plasma with the ambient gas \citep{Missagliaetal,Ternietal,Dehghanetal}.

\textit{Chandra}'s high angular resolution has enabled us to study interesting hydrodynamic phenomena in clusters, e.g., bow shocks driven by the infalling subclusters, cold fronts, or sharp contact discontinuities between regions of gas with different densities and temperatures.
The cold fronts seen as surface brightness edges in X-ray images are caused by the motion of cool, dense gas clouds in the ambient gas, which are either remnants of the infalling subclusters or the displaced gas from the cluster's own cores \citep{MaximAlexey}.
Hence, radio and X-ray observations of clusters of galaxies present the interplay between the radio galaxies along with the extended radio structures and the ICM.
Several important effects of the central radio source on the thermal state of a cluster have been predicted.
One such effect is the Kelvin-Helmholtz instability, which is
likely to develop at a boundary, e.g. the surface brightness edge, and
a straight, collimated radio jet of the moving radio galaxy apparently
flares up when it crosses the edge \citep{RosenHardee2000,Lokenetal1995,Livioetal1980}.
Therefore, very fast, low density, radio jets can remain coherent as they propagate to a few hundred times the jet radii, and as they encounter an X-ray surface brightness edge, they inflate and can be strongly over-pressured relative to the external medium \citep[e.g., NGC\,6251;][]{2005MNRAS.359..363E,Hardcastleteal2002}. 

The prototype, the Coma cluster of galaxies is optically dominated by two giant elliptical galaxies, NGC\,4874, the brightest cluster galaxy, and NGC\,4869 \citep[$z$ = 0.02288;][]{2004AJ....128.1558S}. 
The tailed radio source associated with NGC\,4869 lies near the Coma cluster center, at $\sim$4$^{\prime}$ (= 111~kpc) from NGC\,4874 and is completely embedded within the diffuse radio halo source, together called as Coma-C \citep{Willson1970}.
The source was mapped with the Very Large Array and the extended structure was reported by \citet{Ferettietal1990} and \citet{Dallacasaetal1989}.

\begin{table*}
\caption{The observations}
\begin{center}
\begin{tabular}{clcccccc}
\hline\hline
Band & Obs. ID  &  Obs. Date & $\nu$ & $\Delta\nu$ & t$_{\rm int.}$ & FWHM & \textsc{rms}  \\
   & &       &  (MHz) &   (MHz) & (hour) & ($^{\prime\prime}\times^{\prime\prime}, ^{\circ}$)& ($\mu$Jy~beam$^{-1}$)              \\
 (1) &  \multicolumn{1}{c}{(2)} & (3) & (4) & (5) & (6) & (7) & (8) \\
\hline\noalign{\smallskip}
 \multicolumn{8}{l}{(legacy) GMRT} \\
150 MHz & 18\_022  & 2010 Apr 19 &  147  &  ~6  &  2.3 &13.08$\times$10.32, 46.88 & 908.6 \\
240 MHz & 09TCA01  & 2006 Feb 06 &  240  &  ~8  &  2.5 &11.81$\times$9.87, ~51.00 & 264.1 \\
325 MHz & 09TCA01  & 2006 Feb 11 &  333  &  16  &  2.6 &~8.76$\times$7.24, ~59.99 & ~44.2 \\
610 MHz & 09TCA01  & 2006 Feb 05 &  618  &  16  &  2.5 &~5.41$\times$4.46, ~70.63 & ~18.4 \\
 \multicolumn{8}{l}{uGMRT} \\
250--500 MHz & ddtB270  & 2017 Apr 28 & ~400  & 200  &  1.8 &~6.65$\times$5.90, ~76.39 & ~21.1 \\
1050--1450 MHz & ddtB270  & 2017 Apr 26 & 1250  & 400  &  1.5 &~2.43$\times$1.95, ~69.51 & ~12.7 \\
\hline\hline
\end{tabular}
\end{center}
\label{tab:obs-log}
\tablecomments{Column~7: The position angle (P.A.) is measured from north and counterclockwise. \\
Column~8: \textsc{rms} noise at the half power point; see also Sec~\ref{sec.morph-spec} and \citet[][submitted]{Lal-submitted} for a discussion.}
\end{table*}

In this work, the second paper in the series--new low-frequency upgraded Giant Metrewave Radio Telescope (uGMRT) observations are presented in the first paper \citep[][submitted]{Lal-submitted}--results for the bent head-tail radio source, NGC\,4869 are presented.   We discuss the detailed radio source morphology and the unique spectral structure, and investigate the effect of Kelvin-Helmholtz instability using the hydrodynamic phenomena in the local X-ray environment with \textit{Chandra} data.
The paper is organized as follows.
Our data are summarized in Sec.~\ref{sec.data-red}, followed by a description of the source morphology (Sec.~\ref{sec.morph-spec}), spectral structure (Sec.~\ref{sec:spectra}), and its local hot gas environment (Sec.~\ref{sec:xray-opt}). In Sec.~\ref{sec.discuss},
we examine the source structure, jet bending conditions along with physical conditions within the tail, and
discuss the role of Kelvin-Helmholtz instability at the interface between two gas media which are at different densities and temperatures.
Sec.~\ref{sec.sum-conc} summarizes our conclusions.

Throughout this paper, we adopted a $\Lambda$CDM cosmology with $H_0$ = 70 km s$^{-1}$ Mpc$^{-1}$, $\Omega_{\rm m}$ = 0.27, and $\Omega_{\Lambda}$ = 0.73.
At the redshift of the NGC\,4869, 1 arcsec corresponds to 462~pc at the luminosity distance of 99.8 Mpc.
We define the spectral index, $\alpha$, as $S_\nu$ $\propto$ $\nu^\alpha$, where $S_\nu$ is the flux density at the frequency, $\nu$.
Throughout positions are given in J2000 coordinates.

\section{Radio and X-ray data}
\label{sec.data-red}

The GMRT \citep{Swarupetal1991} has been upgraded with a completely new set of receivers at
frequencies, $<$~1.5 GHz.  The uGMRT now has (nearly) seamless frequency coverage in the 0.050--1.50 GHz range \citep{Guptaetal2017}.

Although the Coma cluster is among the best-studied rich cluster at low frequencies ($<$ 1~GHz), much remains to be explored; however, because of the difficulties encountered with ionospheric refraction and terrestrial radio frequency interference.
It was re-observed in 2017 with the uGMRT at band-3, a 250--500 MHz band
and band-5, a 1050--1450 MHz band, because of much improved sensitivity
\citep[][submitted]{Lal-submitted}.
The archival data and the new observations are detailed in observations log
(Table~\ref{tab:obs-log}).
Both the archival GMRT data and the new uGMRT data were analyzed in \textsc{aips} and \textsc{casa}, using standard imaging procedures \citep[see also][submitted]{LalandRao2004,Lal-submitted}.
Briefly, the flux density calibrator, 3C\,286 was also used as the secondary phase calibrator during the observations.  We used 3C\,286 to correct for the bandpass shape and to set the flux density scale \citep{PerleyButler}.  Bad data and data affected due to radio frequency interference were identified and flagged.  Next the central channels were averaged to reduce data volume, taking care to avoid bandwidth smearing and the visibilities were imaged using the \textsc{tclean} task in \textsc{casa}.
We used 3D imaging (gridder = `widefield'), two Taylor coefficients (nterms = 2), and Briggs weighting (robust = 0.5) in the task \textsc{tclean}.
Several repeated steps of self-calibration, flagging, and imaging were performed to obtain the final image, and the final image was corrected for the primary beam shape of the GMRT antennas.
The error in the estimated flux density, both due to calibration and due to systematics, is $\lesssim$5\%.

\begin{figure*}[ht]
\begin{center}
\begin{tabular}{c}
\includegraphics[width=16.8cm]{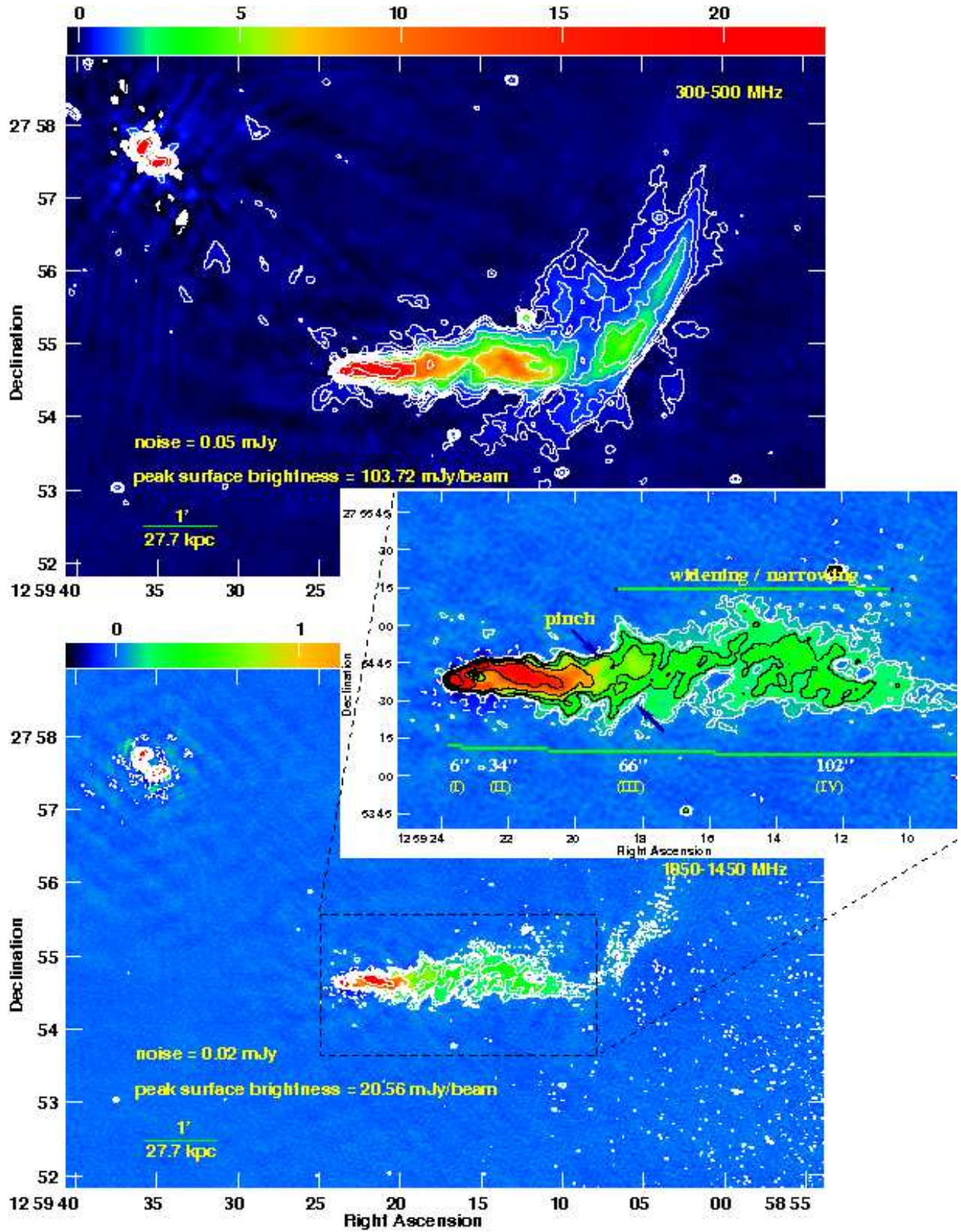} \\
\end{tabular}
\end{center}
\caption{Image of NGC\,4869 at the 250--500 MHz band (upper-panel) and 1050--1450 MHz band (lower panel) of the uGMRT. The double lobed radio emission from NGC\,4874, the brightest cluster galaxy of the Coma cluster, is shown in the northeast corner of the image \cite[R.A.: 12:59:35.71, Dec.: $+$27:57:33.37;][]{SunVFetal2005}. The center frequency is 400 MHz and 1250 MHz at the two bands.  The synthesized beams are 6\farcs65 $\times$ 5\farcs90 at a P.A. of 76\fdg39 and 2\farcs43 $\times$ 1\farcs95 at a P.A. of 69\fdg51, and the \textsc{rms} noise values are 21.1 $\mu$Jy~beam$^{-1}$ and 12.7 $\mu$Jy~beam$^{-1}$ at the half-power points at the 250--500 MHz and 1050--1450 MHz bands, respectively.  The lowest radio contour plotted is three times the local \textsc{rms} noise and increasing by a factors of 2.  The color bar shows the surface brightness, in mJy~beam$^{-1}$. The local \textsc{rms} noise and continuum peak surface brightness of the source are denoted in each panel (lower left corner).  Note that the first two surface brightness contours are not displayed for NGC\,4874 to show the detailed morphology of the head-tail radio source NGC\,4869. The inset illustrates four distinct regions, pinch (marked with arrows), and widening and narrowing of the radio tail (see also Sec.~\ref{sec.morph-spec}).}
\label{fig:f1}
\end{figure*}

\begin{figure*}
\begin{center}
\begin{tabular}{c}
\includegraphics[width=17.6cm]{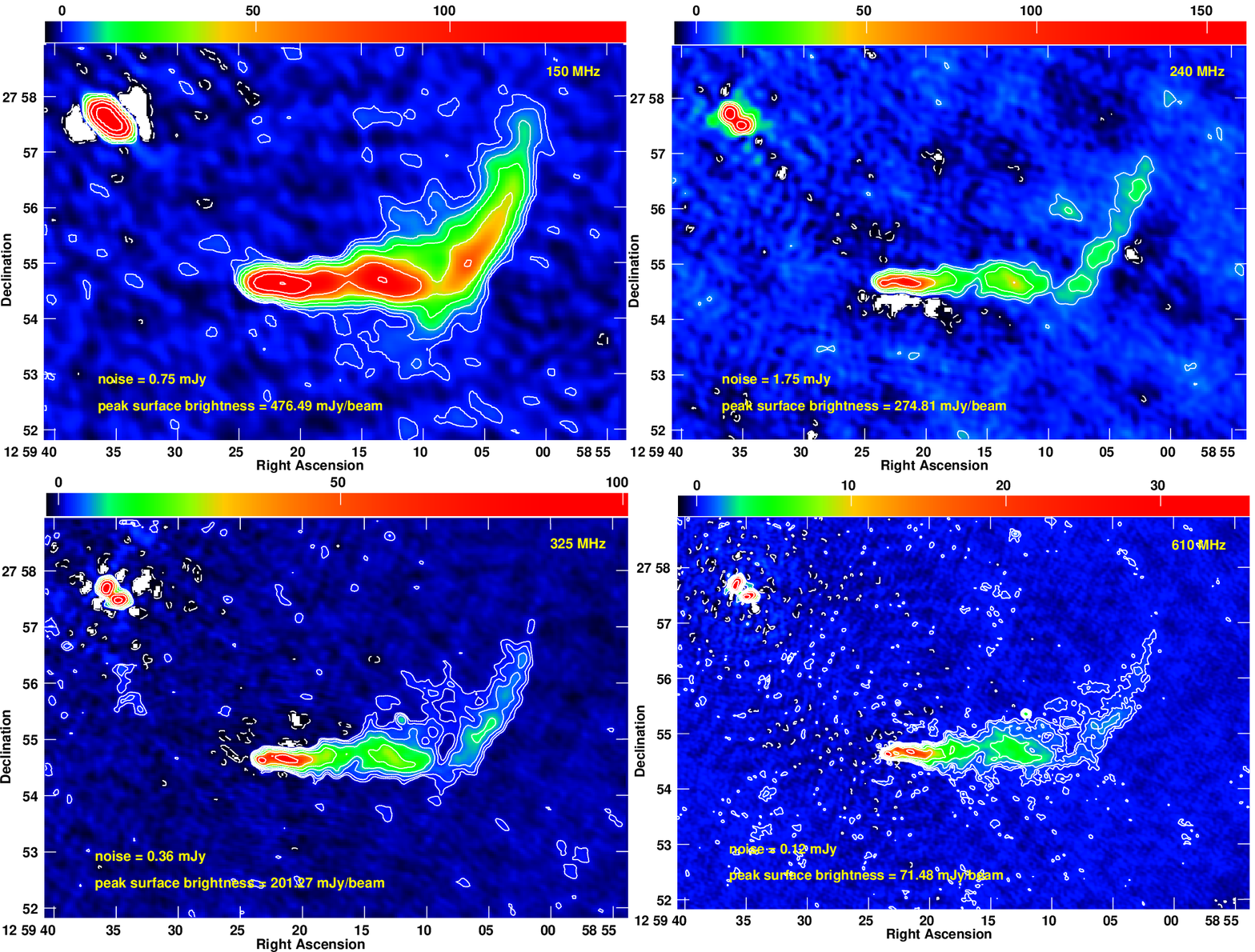}
\end{tabular}
\caption{Gallery of four radio images at 150 MHz (top left), 240 MHz (top right), 325 MHz (bottom left), and 610 MHz (bottom right) using the archival legacy GMRT data.   The synthesized beams and the the \textsc{rms} noise values at the half power points are tabulated in Table~\ref{tab:obs-log}, column 7 and column 8, respectively.  Here again, the lowest radio contour plotted is three times the local \textsc{rms} noise and increasing by factors of 2, and the local \textsc{rms} noise and continuum peak surface brightness of the source are denoted in each panel (lower left corner).  The color bar shows the surface brightness, in mJy~beam$^{-1}$.}
\label{fig:f2}
\end{center}
\end{figure*}

\subsection{Chandra data}
\label{chandra-data}

The Coma cluster has been observed a multiple number of times with \textit{Chandra}.  We used only two observations made in 2012 March, AO-13 and AO-14, which observed NGC\,4869 giving a total observation time of $\sim$115~ks. The \textit{Chandra} observations used are listed here by Obs. IDs 13994 and 14411. These archival data sets were reduced using \textsc{ciao} 4.7 with the most up-to-date gain and efficiency calibrations as of that release using the standard pipeline of \textsc{ciao}.  The standard screening, good time intervals, removal of bad pixels, and grade filtering were applied.  The event data were projected to a common reference point and merged to create images.  All imaging analyses were performed on the combined data set for comparison with our high-sensitivity radio image.  Background-subtracted, exposure-corrected images, using (i) the native resolution of the \textit{Chandra} CCDs, (ii) 4 $\times$ 4 pixel binning, and (iii) 8 $\times$ 8 pixel binning were created in four (soft: 0.5--1.2~keV, medium: 1.2--2.0~keV, hard: 2.0--7.0~keV, and broad: 0.5--7.0~keV) energy bands.
The soft energy band image at 8 $\times$ 8 pixel binning (nearly) matches the angular resolutions of our uGMRT 250--500 MHz band and 1050--1450 MHz band images.  It also showed sufficient contrast in the local hot gas environment where the head-tail radio galaxy NGC\,4869 resides in the Coma cluster.  Hence, we used this soft energy band X-ray image for our further analyses, included to make detailed comparisons with our uGMRT images.

NGC\,4869 is at the edge of the field of view in two \textit{Chandra} pointings and we are interested in relative properties of two regions of interest, which encompass this tailed radio source (see Sec.~\ref{sec:jet-interact}).
The point sources and the central AGN were excised from the data, and the source, background spectra, and the response files were made separately for each \textit{Chandra} observation using the \textsc{specextract} task in \textsc{ciao}.
We then coadded the spectra and averaged the response files for each region.  The data was binned to 10 counts per bin, fitted  with \textsc{xspec} \citep[v12.10.1,][]{Arnaud1996} using isothermal \textsc{apec} model, and we determined a single temperature in the 0.5--5.0~keV energy range within the region of interest.  The absorption column density was set to the Galactic foreground value, $N_{\rm H}$ = 0.9 $\times$ 10$^{20}$ cm$^{-2}$ \citep{DickeyLockman}, the abundances fixed at 0.5 $\times$ solar \citep{Vikhlininetal} with solar abundance ratios of \citet{GrevesseSauval}, and the temperature was free to vary. The $\chi^2$ values (fit statistics at 90\% confidence limits) are 298.2 and 256.4 for 305 degrees of freedom in the two regions of interest.

\section{Results}

\subsection{Radio images}
\label{sec.morph-spec}

The large-scale ($\sim$200~kpc) radio morphology of NGC\,4869 at the 250--500 MHz band and 1050--1450 MHz band of the uGMRT is shown in Figure~\ref{fig:f1}.
The images have \textsc{rms} noise values of 21.1~$\mu$Jy~beam$^{-1}$ and 12.7~$\mu$Jy~beam$^{-1}$ at the half-power points and the dynamic ranges of $\approx$~5300 and $\approx$~1700, respectively.
A gallery of four radio images at 150 MHz, 240 MHz, 325 MHz, and 610 MHz using the archival legacy GMRT data is shown in Figure~\ref{fig:f2}.
The \textsc{rms} noise values and dynamic ranges in archival legacy images are in the ranges 18.4--908.6~$\mu$Jy~beam$^{-1}$ and 90--600, respectively.
The \textsc{rms} noise is a factor of $\approx$2 higher close to the phase center where two dominant and extended radio sources, NGC\,4874, the bright cluster galaxy, and NGC\,4869 are present.
We thus also quote the local \textsc{rms} noise values in both Figure~\ref{fig:f1} and Figure~\ref{fig:f2} (lower-left corner) in all radio images.

These high resolutions and high-sensitivity images show the detailed structure of the radio core, the collimated radio jet, and the tail of NGC\,4869.
The structure is typical of a head-tail radio source: a weak core, two oppositely directed radio jets, and a long, low surface brightness tail.  The tail, which consists of radio-emitting plasma left behind from the galaxy motion, begins after sharp bends in the radio jet.  The low surface brightness, diffuse radio tail after $\sim$210$^{\prime\prime}$ (= 97.0~kpc) is resolved out in our 2\farcs18 image at the 1050--1450 MHz band.  Close inspection of the outer isophotes of the 250--500 MHz band and the 1050--1450 MHz band images show that the tailed jet can be divided into five regions as follows.
\begin{itemize}
\item[(i)] Head (0--6$^{\prime\prime}$ = 0--2.8~kpc): an unresolved radio head.  The two oppositely directed radio jets emanating from the apex of the host galaxy, initially traverse toward northeast and southwest directions.
\item[(ii)] Inner (6$^{\prime\prime}$--40$^{\prime\prime}$ = 2.8--18.5~kpc): a conical shaped feature centered on the nucleus.  As the galaxy plows through the dense intracluster gas, these jets traversing in opposite directions form a trail, after sharp bends in the jets, behind the host galaxy due to interaction with the ICM.
\item[(iii)] Intermediate (40$^{\prime\prime}$--106$^{\prime\prime}$ = 18.5--49.0~kpc): a region in which the jet initially expands much more rapidly and then recollimates.
A pinch at $\approx$1\farcm4 (= 38.8~kpc) is present, which was also reported by \citet{Ferettietal1990}.
\item[(iv)] Outer-flaring (106$^{\prime\prime}$--208$^{\prime\prime}$ = 49.0--96.1~kpc): a second region of expansion.

\begin{figure}[b]
\begin{center}
\begin{tabular}{c}
\includegraphics[width=8.4cm]{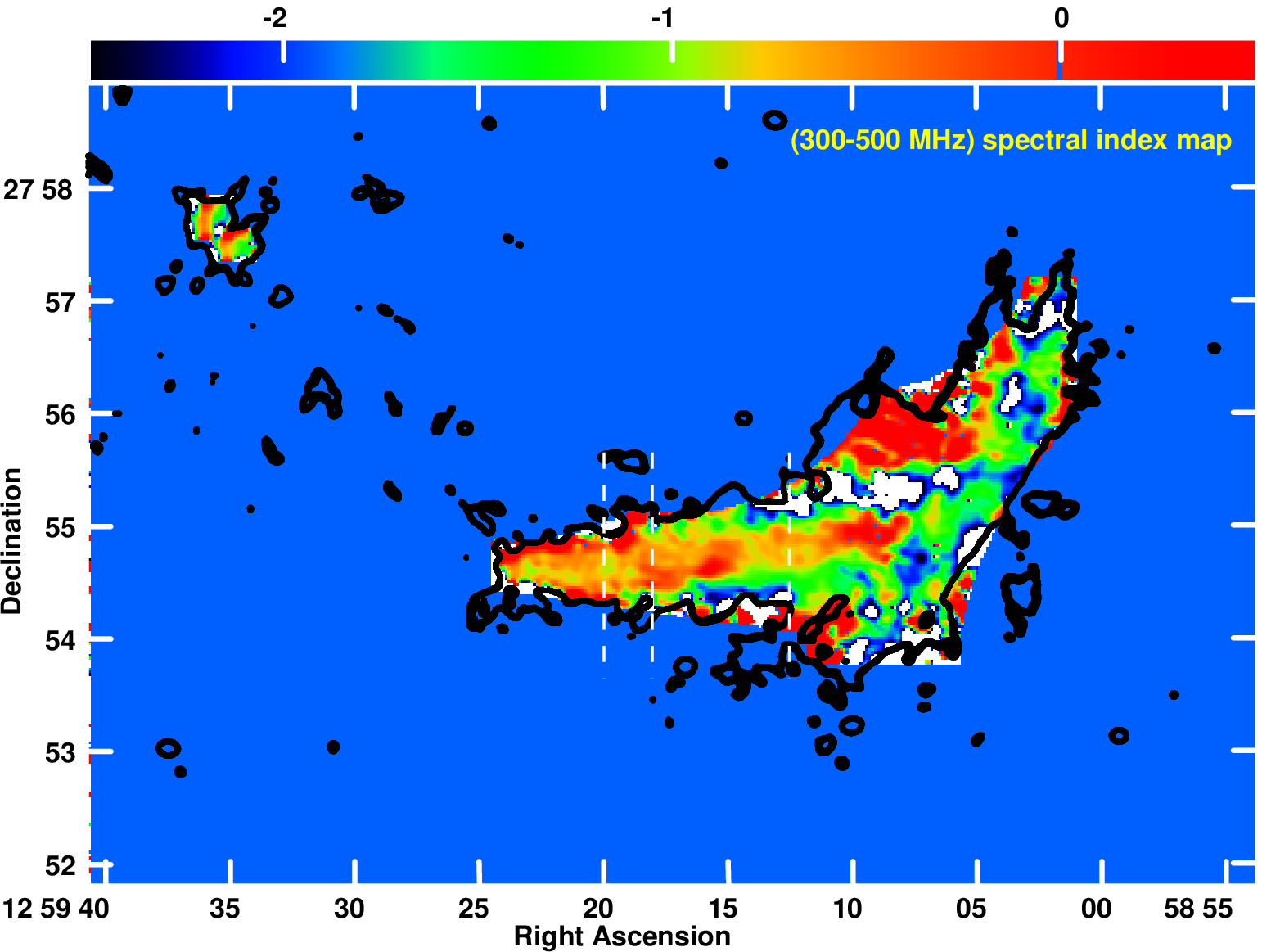} \\
\includegraphics[width=8.4cm]{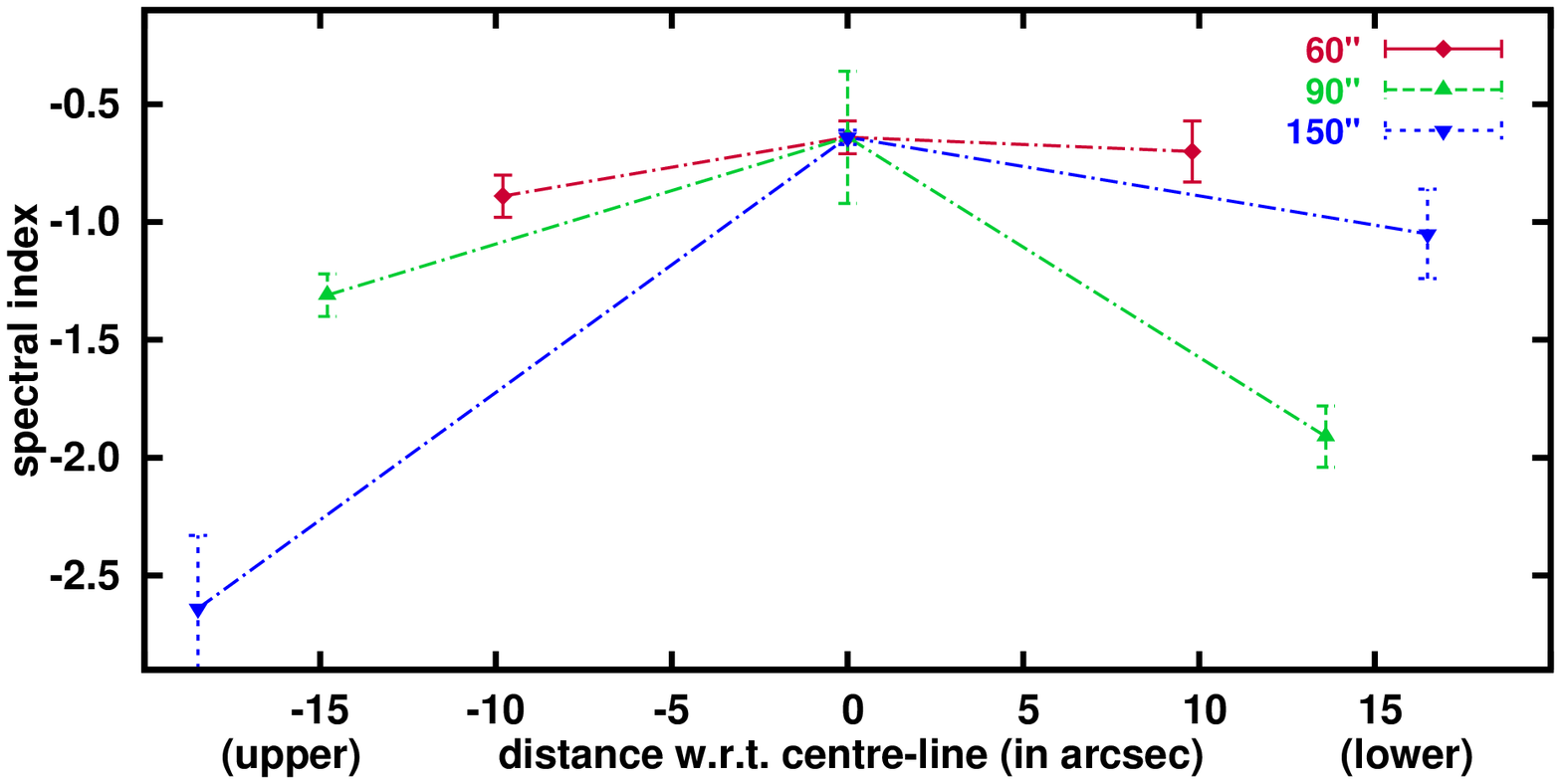}
\end{tabular}
\caption{In-band spectral index image showing spectra using 250--500 MHz band data of uGMRT, between 300 MHz and 500 MHz (upper panel).  The total intensity surface brightness contour is at 0.11 mJy~beam$^{-1}$ from the 300--500 MHz radio image (Figure~\ref{fig:f1}).
The lower panel shows the spectra at three regions along three vertical cuts, at 60$^{\prime\prime}$ (= 27.7~kpc), 90$^{\prime\prime}$ (= 41.6~kpc) and 150$^{\prime\prime}$ (= 69.3~kpc) from the radio head.
The distances on the horizontal axis are the distances from the radio jet ridge line to the region at the upper edge (left) and at the lower edge (right) of the radio jet, where the spectra are determined.
The error bars are based on local \textsc{rms} noise as evaluated in a circle of 2~arcmin diameter centered on these regions.
The locations of these three vertical cuts (white dashed lines) are overlaid, corresponding to the spectra shown in the lower panel figure, in the in-band spectral index image.
The three spectra are shifted (spectra at 60$^{\prime\prime}$, 90$^{\prime\prime}$ and 150$^{\prime\prime}$ cuts are shifted by 0.00, $-$0.18 and $-$0.06, respectively) to match the spectra at center jet ridge line to show comparisons.}
\label{fig:f3}
\end{center}
\end{figure}

\item[(v)] Bent-tail (208$^{\prime\prime}$--312$^{\prime\prime}$ = 96.1--144.1~kpc) and beyond: a feature seen after the jet has flared and bent toward the north.
There is a clear presence of a ridge at $\approx$3\farcm4 (= 94.2~kpc) of radio emission perpendicular to the direction of the tail at the flaring point.  The radio jet in this region is inclined by $\sim$70$^\circ$ with respect to the radio-tail direction.
\end{itemize}

The low-frequency ($\nu$ $<$ 1.0~GHz) legacy GMRT images (Figure~\ref{fig:f2}) also show clear presence of these five regions.
The tail at the beginning has a width of $\sim$6$^{\prime\prime}$ (= 2.8~kpc). It has a constant transverse size of $\sim$16$^{\prime\prime}$ (= 7.4~kpc) at the beginning of the inner region and becomes a factor of two larger close to the pinch, located at $\approx$1\farcm4 (= 38.8~kpc) from the head.  The radio jet widens and narrows several times in the tail,
which continues into the outer flaring region and the radio jet has become weaker.  The bent tail, which reaches out to 6$^{\prime}$ (= 166.3~kpc) and more has nearly an average size of $\sim$26$^{\prime\prime}$ (= 12.0~kpc).
The transverse size of the radio tail is smallest at the pinch, which is seen clearly in 240\,MHz, 610\,MHz, and 1050--1450 MHz band images and is the largest at the ridge, which is seen in 150 MHz, 250--500 MHz, and 325 MHz band images.

The center ridge line of the radio jet shows the signature of oscillations in the jet and all along in the low surface brightness structures in all of our radio images (Figures~\ref{fig:f1} and \ref{fig:f2}), which was reported earlier by \citet{Ferettietal1990} and was seen in the tails of 3C\,129 \citep{LalandRao2004} as well.
The images also show the presence of diffuse extensions toward NGC\,4874 (inner edge) and sharpness in surface brightness on to the outer edge.

\subsection{Radio spectra}
\label{sec:spectra}

\begin{figure}[t]
\begin{center}
\begin{tabular}{c}
\includegraphics[angle=-90,width=8.4cm]{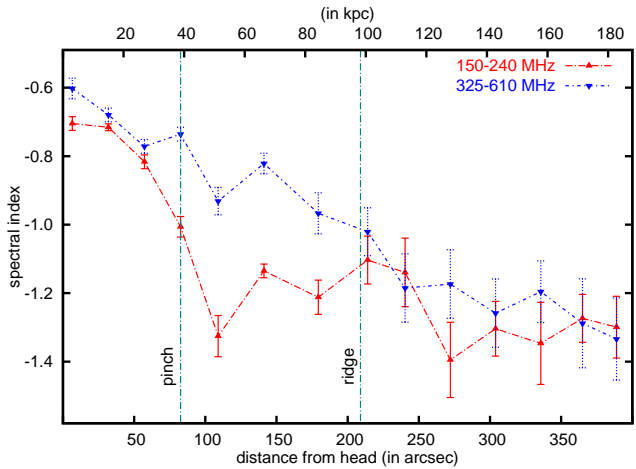}
\end{tabular}
\caption{The plot shows the radio spectra between 150 MHz and 240 MHz (red upper triangles) and between 325 MHz and 610 MHz (blue lower triangles).  The locations of the pinch and the ridge are shown as vertical lines.}
\label{fig:f4}
\end{center}
\end{figure}

The low-frequency data, from both the uGMRT and the legacy GMRT, allow us to derive source spectra at different locations.  Figure~\ref{fig:f3} shows the in-band radio spectral index image between 300 MHz and 500 MHz using 250--500 MHz band data (upper-panel), and
the transverse spectra at three different regions for each of the three different locations, 60$^{\prime\prime}$ (= 27.7~kpc), 90$^{\prime\prime}$ (= 41.6~kpc) and 150$^{\prime\prime}$ (= 63.9~kpc) from the head (lower panel).
Figure~\ref{fig:f4} shows the radio spectra at increasing distance from the host galaxy using matched angular resolution radio images at 150 MHz, 240 MHz, 325 MHz, and 610 MHz.

The upper panel of Figure~\ref{fig:f3} shows progressive spectral steepening as a function of distance from the head.  The spectral indices close to the head and at the end of the outer flaring region are $-$0.66 $\pm$0.04 and $-$1.32 $\pm$0.08, respectively.  Besides, there is a clear presence of two component spectral indices in the inner, the intermediate, and the outer flaring regions, i.e., there is a clear presence of a spine and an enveloping sheath layer, which have different spectra.  The transverse-cut spectra (Figure~\ref{fig:f3}, lower panel) show that the spectrum of the enveloping sheath (upper and lower) is relatively steeper than the spine.
As we progress to larger and larger distances, i.e. from head to $\approx$1\farcm5 (= 41.6~kpc), the spectra of the sheath layer and the spine steepens due to synchrotron cooling, but at a given distance from the head, the sheath layer has relatively steeper spectral index as compared to the spine.  Once the radio jet flares and is bent, the bent tail and beyond region, this effect diminishes and eventually is indistinguishable.

\begin{figure}[b]
\begin{center}
\begin{tabular}{c}
\includegraphics[width=8.4cm]{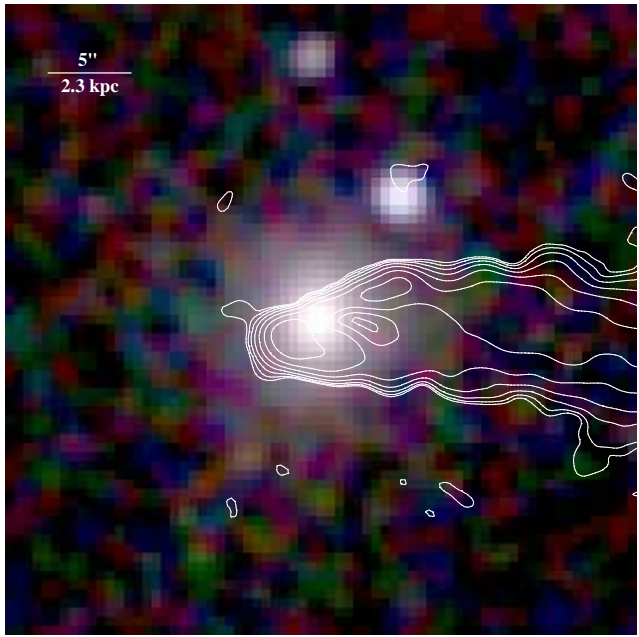}
\end{tabular}
\caption{2MASS, $J$, $H$, and $K_s$ band, composite galaxy image.  The surface brightness contours are displayed from the high-resolution 1050--1450 MHz radio image and the lowest radio contour plotted is three times the local \textsc{rms} noise and increasing by a factor of 2, as shown in Figure~\ref{fig:f1}. The optical image is 1.15 $\times$ 1.15 arcmin$^2$ in size.}
\label{fig:5}
\end{center}
\end{figure}

\begin{figure}[t]
\begin{center}
\begin{tabular}{c}
\includegraphics[width=8.4cm]{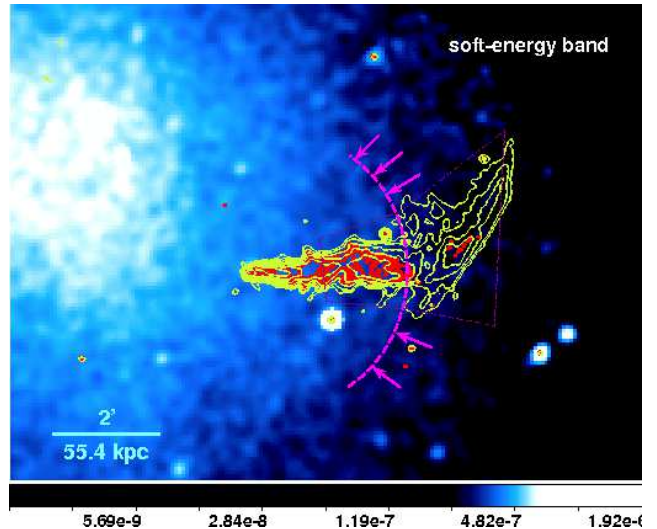}
\end{tabular}
\caption{Background-subtracted and exposure-corrected \textit{Chandra} image at $\sim$4$^{\prime\prime}$ angular resolution in the soft (0.5--1.2 keV) energy band.  The color bar shows the surface brightness, in counts~s$^{-1}$ per 8 $\times$ 8 pixel binning.  The set of red-arrows marks the location of onset of flaring, i.e., the surface brightness edge.  The light-green and red surface brightness contours are displayed using 250--500 MHz band and 1050--1450 MHz band uGMRT images, respectively.  Also shown are the two sectors on either side of the X-ray surface brightness edge that were used for the surface brightness and temperature measurements.}
\label{fig:f6}
\end{center}
\end{figure}

Figure~\ref{fig:f4} shows that the radio spectra at 150--240 MHz and at 325--610 MHz as a function of distance from the head obtained using matched angular resolution images at these frequencies.
The spectrum at 325--610 MHz declines linearly with distance from the radio head (from $-$0.62 $\pm$0.03 and $-$0.77 $\pm$0.04) up to $\sim$60$^{\prime\prime}$ (= 27.7~kpc), whereas the spectrum at 150--240 MHz does not.  Just after the location of the pinch, $\sim$106$^{\prime\prime}$ (= 49.0~kpc) where the intermediate region ends and the outer flaring region begins, there is a dip in spectra showing spectral steepening.  There on, with increasing distance from the head, the radio spectrum shows more steepening over a wide range of frequencies.  
Furthermore, the best fitting regressions to our small range, 150--610 MHz data in the inner, intermediate, and outer flaring regions, show that the high-frequency (325--610 MHz) spectra are flatter than the low-frequency (150--240 MHz) spectra and there on they are indistinguishable.  Quantitatively, the high-frequency (325--610 MHz) spectra changes from $-$0.60 $\pm$0.03 to $-$1.18 $\pm$0.10 and the low-frequency (150--240 MHz) spectra changes from $-$0.70 $\pm$0.02 to $-$1.14 $\pm$0.10.  Whereas, in the bent tail and beyond region, the two spectra are indistinguishable.
Finally, our observed radio spectra at different locations along the radio tail suggest a decrease in the flux density per unit length, but there are no such decrease, e.g., in the inner, the intermediate, and the outer flaring regions in our uGMRT and legacy GMRT images (see Figures~\ref{fig:f1} and \ref{fig:f2}).

\subsection{Optical and X-ray images}
\label{sec:xray-opt}

Figure~\ref{fig:5} shows a Two Micron All Sky Survey (2MASS) optical $J$, $H$, and $K_s$ band composite image.  Overlaid on the optical image are the surface brightness contours from high-resolution, 1050--1450 MHz band image.  The interplay of optical and radio images show that the collimated radio jet expands from 6$^{\prime\prime}$ (= 2.8~kpc) width to 16$^{\prime\prime}$ (= 7.4~kpc) width (transverse size) just outside the transition between the interstellar medium (ISM) of the host galaxy and hot cluster gas \citep{Vikhlininetal}, an effect also seen in NGC\,6109 \citet{Rawesetal2018}.

Figure~\ref{fig:f6} shows a \textit{Chandra} X-ray image at $\sim$4$^{\prime\prime}$ angular resolution in the soft (0.5--1.2 keV) energy band.  Overlaid on the X-ray image are
the surface brightness contours using the 250--500 MHz band image (light-green contours) and the 1050--1450 MHz band image (red contours).
The set of red arrows along with an arc marks the presence of a surface brightness edge.  The ridge of radio emission perpendicular to the direction of the tail coincides with this surface brightness edge.
The radio surface brightness of the collimated jet decreases as soon as the jet crosses the surface brightness edge, which is resolved out in our high-resolution 1050--1450 MHz band image. 

\section{Discussion}
\label{sec.discuss}

The head-tail radio sources are characterized by a highly elongated radio structure.
In the case of NGC\,4869, the source morphology strongly suggests a radio trail, a typical model for head-tail radio sources \citep{JaffePerola1973}.
The two radio jets emanating from the apex of the host galaxy initially traverse along the northeast and southwest as is seen in the head region.  Further on, in the inner region, as the galaxy plows through the dense intracluster gas, these jets traversing in opposite directions form a trail behind the host galaxy due to interaction with the ICM forming a conical shaped feature centered on the nucleus.
The pinch in the intermediate region
is possibly due to (i) changes of the external conditions, or (ii) variations in the nuclear activity, or (iii) the overlap, twist, and wrap of the two tails of the source along with the projection effects.  The tail after the pinch is likely to be at rest with respect to the ambient hot gas \citep{Ferettietal1990}, because the tail consists of blobs of plasma left behind from the moving host galaxy.  But it is unlikely to be in pressure equilibrium with the local X-ray hot gas environment because it does not show a constant width; instead, there is an expansion taking place along with the mixing/twist of the two tails of the source.
This twist of two trailing radio jets seen in the intermediate and outer flaring regions is causing the widening and narrowing several times in the radio tail seen in radio images (see Figures~\ref{fig:f1} and \ref{fig:f2}).
This further argues that the jets are not disrupted even at the bend, instead a continuous bulk flow of blobs of plasma material is still present.  The projection effects too are important, since the host galaxy velocity does not seem to be perpendicular to the projected radio jets \citep{Ferettietal1990}.
The radio spectrum shows progressive steepening with increasing distance from head, because the radio-emitting synchrotron electrons are expected to be of the older population at greater distances from the head.  This variation of the spectrum with distance presents the evolution of radio source under the effects of radiation losses.
There is an absence of a decrease in the flux density per unit length in the 6$^{\prime\prime}$--208$^{\prime\prime}$ (= 2.8--96.1~kpc) region, which suggests that the re-acceleration of the radiating electrons and perhaps also the magnetic field regeneration is possibly occurring within the jet \citep[e.g., Centaurus\,A;][]{HESScollaboration}.
The diffuse extensions toward NGC\,4874 (inner edge) and sharpness in surface brightness on to the outer edge along with the curvature of the tail of NGC\,4869 suggests a closed orbit around the dark matter potential.
It is not, however possible to estimate the cluster mass needed to produce the observed
curved radio structure and bent trajectory on kiloparsec scales extending away from the active galactic nuclei (AGN).  This is because we lack knowledge of both the tangential velocities and the complete orbits, and the trail of NGC\,4869.  The latter, the trail has the form of an overlap, twist, and wrap of the two tails, which could reflect the rotation of the region of the galaxy that is ejecting radio-emitting blobs of plasma \citep[see also][]{Gendron-Marsolaisetal}.

More scrutiny of the properties of NGC\,4869, however, reveals a difficulty in this simple radio trail model.
The problem centers around the end of the outer flaring region and beginning of bent tail region, where (i) the jet bends by $\sim$70$^\circ$ with respect to initial radio-tail direction and (ii) there is a clear presence of a ridge, flaring of the jet, perpendicular to the propagation of the radio jet, an aspect we discuss below.

\subsection{Interaction of radio tail with the surrounding material}
\label{sec:jet-interact}

The jets in FR\,I radio galaxies decelerate by picking up the matter, i.e., the entrainment across the jet boundary \citep{Bicknell1984, Bicknell1986,deYoung1996}.  This is because the edge of the jet or the sheath layer is traveling about 30\% more slowly than the center, or the spine of the jet \citep{2002MNRAS.336..328L,2002MNRAS.336.1161L}, suggesting evidence of an interaction between the radio jet and the external medium.  \citet{2002MNRAS.336.1161L} also showed that the radio jets decelerate by entrainment and are recollimated by the external pressure gradient, which is also quantitatively in agreement with the typical velocity field for external gas parameters derived from \textit{Chandra} measurements, \citep[e.g., FR\,I radio galaxy 3C\,31;][]{Hardcastleteal2002}.
The emission line data for several 3C\,RR radio galaxies imply masses of 10$^7$ M$_\sun$ in the line emitting gases.  If a large part of this gas is entrained, then a radio jet could accomplish this in 10$^6$--10$^7$ yr \citep{deYoung1986}.
The entrained mass of $\sim$10$^8$ M$_\sun$ or more is expected for NGC\,4869, with a radio tail of radius $\sim$1~kpc, velocities around 1000 km~s$^{-1}$ during the $\simeq$10$^8$ yr lifetime of the radio jet. Therefore, in NGC\,4869, it is clear that a boundary layer must form due to entrainment, which transfers momentum to the ambient gas.
The radio spectra, presented in Sec.~\ref{sec:spectra} show that the
enveloping sheath is of steeper spectra than the spine, which we suggest is due to the entrained material from the ICM.
As we progress to larger and larger distances from the head, the sheath layer has entrained more and more thermal gas material and becomes steeper and steeper as compared to the spine at a given location in the radio tail.

The surface brightness edge (Figure~\ref{fig:f6}), where the sudden transition occurs or the jet abruptly flares represents the onset of turbulence \citep{Bicknell1984} or, in other words, the point at which Kelvin-Helmholtz instabilities start to grow nonlinearly \citep{RosenHardee2000,Lokenetal1995}.
The instability under consideration is instability of the boundary, also called the surface brightness edge between the two gas media, which are at different densities and temperatures.
Both the boundary between the ISM of the moving galaxy and the surrounding ICM, and the surface brightness edge seen in the hot gas media exhibit a large drag to the propagating, straight, collimated radio jet causing it to flare.
In the former, the size of the ISM is small ($\sim$5.5~kpc, NGC\,4869) and the amount of ISM in the galaxy is also small \citep{Vikhlininetal}.  Its effect is stripping of gas in the ISM due to the motion of the host galaxy \citep{McBrideMcCourt,Livioetal1980}.
Whereas in the latter, jets crossing density edges can become partially disrupted and they inflate as a consequence of interactions between an AGN and its surrounding medium \citep[e.g., 3C\,449;][]{Laletal2013}.

\begin{figure}[t]
\begin{center}
\begin{tabular}{c}
\includegraphics[width=8.4cm]{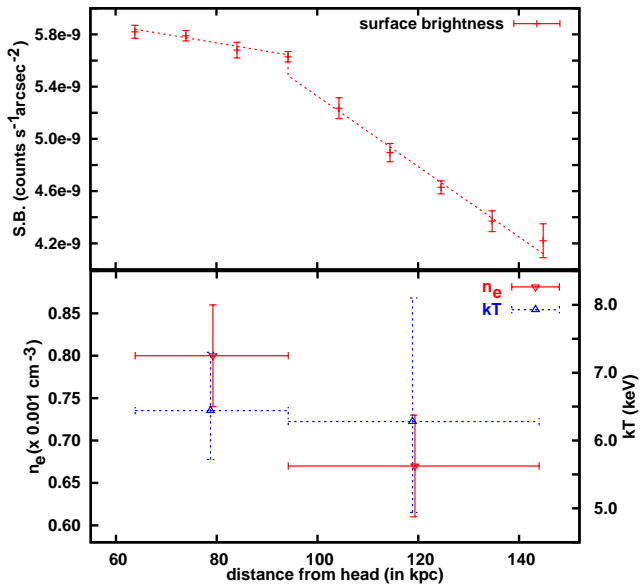}
\end{tabular}
\caption{Plot of the radial surface brightness (upper panel), and the gas density and the temperature (lower panel). The error bars are at 90\% confidence intervals.}
\label{fig:f7ab}
\end{center}
\end{figure}

The central AGNs can have significant impacts on the small ISM of the host galaxy through its jets, especially in outbursts \citep[e.g., NGC\,6109;][]{Rawesetal2018}.
The jet power, which is distributed in relativistic electrons, heavier particles, and the magnetic field, is known to be much larger than the radio luminosity of the source.
Following \citet{Sunetal2005}, we estimate the kinetic power of jets as $\simeq$10$^{43}$ ergs~s$^{-1}$.  The (minimum) energy in the magnetic field (see also Sec.~\ref{sec.source-conf}) is only 10\%$-$25\% of the kinetic power and the AGN has been active for $>$10$^8$~yr.  Therefore, the total jet power over the active phase of the nucleus is at least $\gtrsim$1000 times the thermal energy in the ISM.  In other words, the jets carry large amounts of energy without dissipation through the small ISM of the host galaxy.
Instead, as reported in Sec~\ref{sec.morph-spec}, in the inner region, a cone shaped feature expands to form a jet of $\sim$16$^{\prime\prime}$ (= 7.4~kpc) width (transverse size) as they leave the small ISM of the host galaxy into the cluster atmosphere, possibly because of the change in the external pressure.
Fortunately, the host galaxy retains most of its ISM and only a small fraction is stripped \citep[e.g., NGC\,4848 and IC\,4040 in Coma cluster;][]{Chenetal}

Next, we focus our discussion on the latter instability, in the outer flaring region, which we have detected in our observations and seen in the \textit{Chandra} X-ray image (Figure~\ref{fig:f6}).
Observations show that the apparent pressures within many jets are higher than the pressures in the surrounding X-ray emitting gas through which they propagate e.g., NGC\,6251; \citet{2005MNRAS.359..363E} and 3C31; \citet{2002MNRAS.336..328L,2002MNRAS.336.1161L}.
Additionally, the X-ray emission shows that the Coma cluster is unrelaxed and has undergone a recent cluster merger \citep{Arnaudetal2001,Brieletal2001}, and the surface brightness edge (see Figure~\ref{fig:f6}) seen is possibly a cold front. Hence, the pressure must be continuous through the interfaces between the two gas medias, and a density and temperature jumps are only expected.
We observed the X-ray edge, where the surface brightness drops, and we select two regions inside and outside this edge (see Figure~\ref{fig:f6}).  Since the plasma X-ray emissivity is proportional to the square of the gas density, the surface brightness distribution provides a direct estimate of the gas density \citep{Johnsonetal}.
Indeed, our surface brightness shows a discontinuity at the edge, and therefore implies a discontinuity, or jump, in the gas density (Figure~\ref{fig:f7ab}, upper panel).
We model the density within the two regions, following \citet{Johnsonetal}, with a power-law function and the density measurements are $n_e$(inside) and $n_e$(outside) $\approx$ 0.80 $\pm$0.06 $\times$ 10$^{-3}$ cm$^{-3}$ and 0.67 $\pm$0.06 $\times$ 10$^{-3}$ cm$^{-3}$, respectively (Figure~\ref{fig:f7ab}, lower panel).
We extracted source spectra for two regions using isothermal model, inside and outside of the surface brightness edge encompassing the source (see also Sec.~\ref{chandra-data} and Figure~\ref{fig:f7ab}, lower panel).
The derived temperatures are $\approx$ 6.44$^{+0.86}_{-0.72}$~keV and 6.28$^{+1.82}_{-1.34}$~keV for $kT_{\rm (inside)}$ and $kT_{\rm (outside)}$, respectively.  Although with large error bars, these temperature measurements are hinting that the surface brightness edge is possibly a cold front.  They are also in reasonable agreement with coarser measurements of \citet[][Figure~5]{Arnaudetal2001} and \citet[][Figure~3]{Brieletal2001}.
Furthermore, the two gas medias retain their respective material despite the motion of the radio galaxy, and conversely, the radio-emitting plasma too is not immune to stripping for every crossing of the surface brightness edge \citep{McBrideMcCourt,2013Sci...341.1365S}.
It, therefore, seems that the ram pressure condition indicates stability against the ablation of radio emitting plasma; in other words, the Kelvin-Helmholtz instability will induce a flaring of the collimated radio jet.  In view of the vulnerability of the collimated jet at the location of a surface brightness edge, we conclude the onset of flaring, seen in several FR\,I radio galaxies \citep[3C\,449;][]{Laletal2013} and also seen here in the soft energy (0.5--1.2 keV) band \textit{Chandra} X-ray image (Figure~\ref{fig:f6}) of NGC\,4869.
Our conclusion is supported by the hydrodynamic simulations \citep{Lokenetal1996,Lokenetal1995} of the motion of radio galaxies through the hot intracluster gas, which shows that a Kelvin-Helmholtz instability manifests itself very prominently in the form of dense and long tongues of material at the interface between the moving radio source and the surface brightness edge in the gas.
Finally, the jet is bent by $\sim$70$^{\circ}$ toward the north direction; since the host galaxy could not be moving perpendicular to the trailing jets, the projection of the source is a plausible explanation for this bend \citep[also see NGC\,6109;][]{Rawesetal2018}.

\subsection{External density, pressure, and radiative age}
\label{sec.source-conf}

\begin{table}[t]
\caption{Physical parameters of NGC\,4869}
\begin{center}
\begin{tabular}{cccccc}
\hline
Dist. &  Size & $U_{\rm min}$   & $B_{\rm eq.}$ & $P_{\rm min}$ & t$_{\rm age}$ \\
        & & $\times$ 10$^{-12}$ & & $\times$ 10$^{-12}$ & 10$^{8}$ \\
($\prime\prime$) & (kpc) & (erg~cm$^{-3}$) & ($\mu$G) & (dyne~cm$^{-2}$) & (yr) \\
   \multicolumn{1}{c}{(1)} & (2) & (3) & (4) & (5) & (6) \\
\hline\noalign{\smallskip}
 ~~3.1 & 10.4 & 3.9 & 6.3 & 1.3 & 1.2 \\
 ~15.0 & 10.4 & 3.8 & 6.2 & 1.3 & 1.1 \\
 ~26.9 & 10.4 & 3.1 & 5.7 & 1.0 & 1.2 \\
 ~38.8 & 12.2 & 2.0 & 4.6 & 0.7 & 1.7 \\
 ~51.2 & 12.2 & 1.5 & 4.0 & 0.5 & 2.1 \\
 ~66.4 & 17.6 & 1.5 & 3.9 & 0.5 & 2.1 \\
 ~84.3 & 17.6 & 1.1 & 3.4 & 0.4 & 2.6 \\
 100.6 & 12.0 & 0.7 & 2.7 & 0.2 & 3.7 \\
 113.0 & 14.7 & 1.0 & 3.2 & 0.3 & 3.0 \\
 127.9 & 14.7 & 1.0 & 3.1 & 0.3 & 3.0 \\
 142.8 & 14.7 & 0.7 & 2.6 & 0.2 & 3.9 \\
 157.8 & 14.7 & 0.6 & 2.5 & 0.2 & 4.1 \\
 171.5 & 12.8 & 0.5 & 2.2 & 0.2 & 5.0 \\
 182.8 & 10.0 & 0.4 & 2.0 & 0.1 & 5.9 \\
\hline
\end{tabular}
\end{center}
\label{tab:phy-param}
\end{table}

\begin{figure}[t]
\begin{center}
\begin{tabular}{c}
\includegraphics[width=8.4cm]{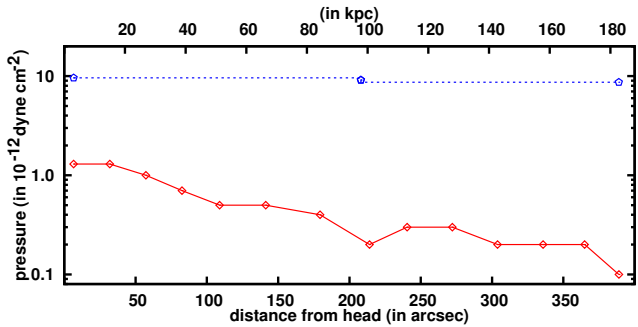}
\end{tabular}
\caption{Plot of the minimum internal non-thermal pressure as a function of distance from head.  The continuous line represents the thermal pressure of the local X-ray gas. The error-bars of the thermal pressure of the local X-ray gas are $\approx$12\%.}
\label{fig:f8}
\end{center}
\end{figure}

The minimum energy condition corresponds almost to the equipartition of energy between relativistic particles and the magnetic field.
In order to determine the minimum internal energy density, equipartition magnetic energy, and pressure for several locations in the radio source, NGC\,4869,
we assume the ratio of energy in the heavier particles and the electrons is unity, the filling factor of the emitting regions also unity, and the transverse size of each component in the source is the path length through the source along the line of sight, and use the standard formulae in \citet{Miley1980} and \citet{Pacholczyk1970}.
Next, the shape of the radio spectrum depends on the balance between
the rate of synchrotron and inverse Compton losses and the rate of the replenishment of the electrons in the radiating region for the power-law energy distribution of electrons.
Following \citet{Miley1980}, we also derive the upper limit to the ages of the synchrotron electrons at several positions along the radio tail.
Table~\ref{tab:phy-param} gives these physical parameters, minimum energy density, equipartition magnetic field, the pressure exerted by the relativistic gas in the radio source, $U_{\rm min}$, $B_{\rm eq.}$, $P_{\rm min}$, respectively, and
radiative age, t$_{\rm age}$ estimated using 240 MHz as the break frequency at different locations along the radio tail.
Figure~\ref{fig:f8} shows that the physical conditions in the source as a function of the distance from the head change, whereas the ambient thermal pressure of the ambient X-ray gas is more and (nearly) constant.

\section{Conclusions}
\label{sec.sum-conc}

In this paper, we have presented in the preceding sections details of the observations and provide a detailed descriptions of the radio morphologies using uGMRT and legacy GMRT data
of an interesting head-tail radio source, NGC\,4869 in the Coma cluster of galaxies.
Our main conclusions are as follows.
\begin{enumerate}
\item[(i)] The high-resolution, high-sensitivity large-scale (almost 200~kpc) structure of the NGC\,4869 source is shown in Figures~\ref{fig:f1} and \ref{fig:f2}. The elliptical host galaxy shows a weak radio core,
two oppositely directed radio jets, and a long--low surface brightness tail.
\item[(ii)] The radio source can be divided into five distinct regions: a head, conical shaped feature centered on the nucleus, a region showing the rapid an expansion of the jet followed by collimation, another region showing expansion of the jet, and a flared region with a bend.  Also present are pinch and a ridge at $\approx$1\farcm4 and $\approx$3\farcm4, respectively, from the head in addition to widening and narrowing several times in the radio tail.  
\item[(iii)] There is no increase in the flux density per unit length in the 6$^{\prime\prime}$--208$^{\prime\prime}$ (= 2.8--96.1~kpc) region, which suggests that the re-acceleration of the radiating electrons and perhaps also magnetic field regeneration could be occurring within the jet.
\item[(iv)] The characteristic feature of radio tail is the sharp bend toward the north at
3\farcm5 from the position of the host galaxy; the bent tail and beyond region is inclined by $\sim$70$^\circ$ with respect to
the radio-tail direction.  It seems that projection effects are important in the radio jets because the host galaxy does not seem to be moving perpendicular to the projected trailing jets.
\item[(v)] The radio spectra show progressive spectral steepening as a function of distance from the radio head due to synchrotron cooling.
The high-frequency (325--610 MHz) spectra are marginally flatter than the low-frequency (150--240 MHz) spectra until the ridge, i.e., in the inner, intermediate, and outer flaring regions.  Once the radio jet flares, the bent tail and beyond region, these spectral features are indistinguishable.
\item[(vi)] Our improved transverse resolution in the 250--500 MHz band data for the inner, the intermediate, and the outer flaring regions clearly show
presence of a steep spectrum sheath enveloping a flat spectrum spine in the in-band spectral index image (Figure~\ref{fig:f3}), hinting at a transverse velocity structure with a fast spine surrounded by a slower sheath layer.
Our results favor entrainment across the boundary layer as the origin of the mass loading of the jets \citep[see also, 3C31;][]{2002MNRAS.336..328L,2002MNRAS.336.1161L}. 
\item[(vii)] We have detected flaring of a straight, collimated radio jet as it crosses a surface brightness edge seen in the \textit{Chandra} X-ray image (Figure~\ref{fig:f6}).  At the flaring point is a surface brightness edge at which the jet collimation, emissivity, and possibly the velocity change abruptly, which is a general property of FR\,I jets \citep{Laingetal1999}.
This sudden transition represents the onset of turbulence \citep{Bicknell1984} or in other words, the point at which Kelvin-Helmholtz instabilities start to grow nonlinearly \citep{RosenHardee2000,Lokenetal1995}.
\item[(viii)] The physical conditions in the jets and along the tails are obtained.  The minimum internal nonthermal pressure is smaller than the external thermal pressure.  This suggests that there is present a significant quantity of thermal plasma within the radio-emitting regions.
\end{enumerate}

Our key results favor (i) entrainment across the boundary layer, (ii) occurrence of re-acceleration of the synchrotron electrons and perhaps also regeneration of the magnetic field within the radio jet, and (iii) flaring of a straight, collimated radio jet as it crosses a surface brightness edge due to Kelvin-Helmholtz instabilities.  Therefore, it will be fruitful to determine the origin and distribution of the slow-moving (sheath layer) material with respect to (spine) material, and the extent to which these regions of FR\,I jets resemble the larger-scale jets in FR\,II sources.  The next step is to seek further evidence for the entrainment process, such as the reduced polarization near the boundaries of the flaring regions using polarization data \citep[e.g., NGC\,6251;][]{2015IAUS..313..108L}, which could also shed light on re-acceleration of the synchrotron electrons and the regeneration of the magnetic field in the radio jets.  Radio jets in a number of FR\,I sources have been detected at X-ray and/or optical wavelengths. The radiation is most plausibly produced by the synchrotron process over the entire observed frequency range, and the shape of the spectrum, therefore, carries information about particle acceleration and energy loss. It will be important to incorporate our uGMRT results into descriptions of these processes for improved models.
It will be equally important to observe FR\,I radio sources in galaxy clusters over a broad range of low frequencies to study in detail the flaring of a collimated radio jet as it crosses a surface brightness edge (usually seen in X-ray images), which are indicators of Kelvin-Helmholtz instabilities.

\smallskip

We thank the anonymous referee for the comments that improved this paper.
D.V.L. would like to thank Tiziana Venturi and W. R. Forman for useful discussions, and Ishwara-Chandra C.H. for discussions on some aspects of this project.
D.V.L. acknowledges the support of the Department of Atomic Energy, Government of India, under project No. 12-R\&D-TFR-5.02-0700.
We thank the staff of the GMRT who made these observations possible. The GMRT is run by the National Centre for Radio Astrophysics of the Tata Institute of Fundamental Research.
This research has made use of the NED, which is operated by the Jet Propulsion Laboratory, Caltech, under contract with the NASA, and NASA's Astrophysics Data System.
This research has made use of the NASA/IPAC Infrared Science Archive, which is funded by the NASA and operated by the California Institute of Technology.
This research has made extensive use of SAOImage DS9, in addition to software provided by the CXC in the application packages CIAO.

Facilities: \facility{Chandra, IRSA, GMRT}


\begin{thebibliography}{}

%
\bibitem[Arnaud et~al.(2001)]{Arnaudetal2001} Arnaud, M., Aghanim, N., Gastaud, R., et~al. 2001, A\&A, 365, L67

\bibitem[Arnaud(1996)]{Arnaud1996} Arnaud K. A., 1996, in Jacoby G. H., Barnes J., eds, ASP Conf. Ser. Vol.101, Astronomical Data Analysis Software and Systems V. Astron. Soc. Pac., San Francisco, p. 17

\bibitem[Bicknell(1986)]{Bicknell1986} Bicknell, G. V. 1986, ApJ, 300, 591

\bibitem[Bicknell(1984)]{Bicknell1984} Bicknell, G. V. 1984, in Physics of energy transport in extragalactic radio sources, eds. A. H. Bridle \& J. A. Eilek (NRAO: Green Bank, WV) p. 229

%
\bibitem[Briel et~al.(2001)]{Brieletal2001} Briel, U. G., Henry, J. P., Lumb, D. H., et~al. 2001, A\&A, 365, L60

\bibitem[Chen et~al.(2020)]{Chenetal} Chen, H., Sun, M., Yagi, M., et~al. 2020, arXiv:2004.0674

\bibitem[Dallacasa et~al.(1989)]{Dallacasaetal1989} Dallacasa, D., Feretti, L., Giovannini, G. \& Venturi, T.  1999, A\&ASS, 79, 391

\bibitem[de~Young(1996)]{deYoung1996} De Young D. S. 1996, in Hardee P. E. Bridle A. H. Zensus J. A., eds, ASP Conf. Series 100, Energy Transport in Radio Galaxies and Quasars, Astron. Soc. Pac., San Francisco, p. 261

%
\bibitem[de~Young(1986)]{deYoung1986} De Young, D. S. 1986, ApJ, 307, 62


\bibitem[Dehghan et~al.(2014)]{Dehghanetal} Dehghan, S., Johnston-Hollitt, M., Franze, T. M. O., et~al. 2014, AJ, 148, 75

\bibitem[Dickey \& Lockman(1990)]{DickeyLockman} Dickey, J. M. \& Lockman, F. J. 1990, ARA\&A, 28, 215

\bibitem[Evans et~al.(2005)]{2005MNRAS.359..363E} Evans, D. A., Hardcastle, M. J., Croston, J. H., et~al. 2005, MNRAS, 359, 363

\bibitem[Feretti et~al.(1990)]{Ferettietal1990} Feretti, L., Dallacasa, D., Giovannini, G., \& Venturi, T. 1990, A\&A, 232, 337

\bibitem[Gendron-Marsolais et~al.(2020)]{Gendron-Marsolaisetal} Gendron-Marsolais, M., Hlavacek-Larrondo, J., van Weeren, R. J., et~al. 2020, arXiv:2005.12298

\bibitem[Gupta et~al.(2017)]{Guptaetal2017} Gupta, Y., Ajithkumar, B., Kale, H., et~al. 2017, Current Science, 113, 707

\bibitem[Grevesse \& Sauval(1998)]{GrevesseSauval} Grevesse, N. \& Sauval, A. J., 1998, Space Sci. Rev., 85, 161

\bibitem[The H.E.S.S. Collaboration(2020)]{HESScollaboration} The H.E.S.S. Collaboration 2020, Nature, 582, 356

\bibitem[Hardcastle et~al.(2002)]{Hardcastleteal2002} Hardcastle, M. J., Worrall, D. M., Birkinshaw, M., et~al. 2002, MNRAS, 334, 182

\bibitem[Jaffe \& Perola(1973)]{JaffePerola1973} Jaffe, W. J. \& Perola, G. C. 1973, A\&A, 26, 423

\bibitem[Johnson et al.(2010)]{Johnsonetal}  Johnson, R. E., Markevitch, M., Wegner, G. A., et~al. 2010, ApJ, 710, 1776

%
\bibitem[Laing \& Bridle(2014)]{2015IAUS..313..108L}
Laing R. A. \& Bridle A. H. 2014, in Massaro, F., Cheung, C. C. \& Lopez A., eds, IAU Symposium, Volume 313, Extragalactic jets from every angle, Proceedings of the International Astronomical Union, p. 313

%
\bibitem[Laing \& Bridle(2002b)]{2002MNRAS.336.1161L} Laing, R. A. \& Bridle, A. H. 2002, MNRAS, 336, 1161

\bibitem[Laing \& Bridle(2002a)]{2002MNRAS.336..328L} Laing, R. A. \& Bridle, A. H. 2002, MNRAS, 336, 328

\bibitem[Laing et~al.(1999)]{Laingetal1999} Laing, R. A., Parma, P., De Ruiter, H. R. \& Fanti, R. 1999, MNRAS, 306, 513

\bibitem[Lal(2020)]{Lal-submitted} Lal, D. V. 2020, ApJS, 250, 22

\bibitem[Lal et~al.(2013)]{Laletal2013} Lal, D. V., Kraft, R. P., Randall, S. W., et~al. 2013, ApJ, 764, 83

\bibitem[Lal \& Rao(2004)]{LalandRao2004} Lal, D. V. \&  Rao, A. P. 2004, A\&A, 420, 491

\bibitem[Livio, Regev \& Shaviv(1980)]{Livioetal1980} Livio, M., Regev, O., \& Shaviv, G. 1980, ApJ, 240, L83

\bibitem[Loken et~al.(1996)]{Lokenetal1996} Loken, C., Burns, J. O., Bryan, G. \& Norman M. 1996, in Hardee P. E. Bridle A. H. Zensus J. A., eds, ASP Conf. Series 100, Energy Transport in Radio Galaxies and Quasars, Astron. Soc. Pac., San Francisco, p. 267

\bibitem[Loken et~al.(1995)]{Lokenetal1995} Loken, C., Roettiger, K., Burns, J. O., \& Norman, M. 1995, ApJ, 445, 80

\bibitem[Markevitch \& Vikhlinin(2007)]{MaximAlexey} Markevitch, M. \& Vikhlinin, A. 2007, Physics Reports, 443, 1

\bibitem[McBride \& McCourt(2014)]{McBrideMcCourt} McBride, J. \& McCourt, M. 2014, MNRAS, 442, 838

\bibitem[Miley(1980)]{Miley1980} Miley, G. K. 1980, ARA\&A, 18, 165

%
\bibitem[Missaglia et~al.(2019)]{Missagliaetal}Missaglia, V., Massaro, F., Capetti, A., et~al. 2019, A\&A, 626, A8
 
%
\bibitem[Pacholczyk(1970)]{Pacholczyk1970}Pacholczyk, A. G. 1970, Radio Astrophysics (San Francisco: W. E. Freeman \& Co.)

\bibitem[Perley \& Butler(2017)]{PerleyButler} Perley, R. A. \& Butler, B. J. 2017, ApJS 230, 7

\bibitem[Rawes, Birkinshaw \& Worrall(2018)]{Rawesetal2018} Rawes, R., Birkinshaw, M. \& Worrall D. M. 2018, MNRAS, 480, 3644

\bibitem[Rosen \& Hardee(2000)]{RosenHardee2000} Rosen, A. \& Hardee, P. E. 2000, ApJ, 542, 750

\bibitem[Sanders et~al.(2013)]{2013Sci...341.1365S} Sanders, J. S., Fabian, A. C., Churazov, E., et~al. 2013, Science, 341, 1365

\bibitem[Smith et~al.(2004)]{2004AJ....128.1558S} Smith, R. J., Hudson, M. J., Nelan, J. E., et~al. 2004, AJ, 128, 1558
\bibitem[Sun, Jerius \& Jones(2005)]{Sunetal2005} Sun, M., Jerius, D. \& Jones, C. 2005, ApJ, 633, 165

\bibitem[Sun et~al.(2005)]{SunVFetal2005} Sun, M., Vikhlinin, A., Forman, W. R., et~al. 2005, ApJ, 619, 169

\bibitem[Swarup et~al.(1991)]{Swarupetal1991} Swarup, G., Ananthakrishnan, S., Kapahi, V. K., et~al. 1991, Current Science, 60, 95

\bibitem[Terni de~Gregory et~al.(2017)]{Ternietal} Terni de~Gregory, B., Feretti, L., Giovannini, G., et~al. 2017, A\&A, 608, A58

\bibitem[Vikhlinin et~al.(2001)]{Vikhlininetal} Vikhlinin, A., Markevitch, M. M., Forman, W. R. \& Jones, C. 2001, ApJ, 555, L87

%
\bibitem[Willson(1970)]{Willson1970} Willson, M. A. G. 1970, MNRAS, 151, 1

\end{thebibliography}
\end{document}